%% file: main.tex
\documentclass[10pt,conference]{IEEEtran}
\usepackage{xcolor}
\usepackage{amsmath} 
\usepackage{graphicx} 
\usepackage{listings}
\usepackage{comment}
\usepackage{url}
\usepackage{hyperref}
\definecolor{tgreen}{rgb}{0,0.6,0}

\begin{document}

\title{QADL: Prototype of Quantum Architecture Description Language}

\author{
    \IEEEauthorblockN{
        Muhammad Waseem\IEEEauthorrefmark{1}, 
        Tommi Mikkonen\IEEEauthorrefmark{1}, 
        Aakash Ahmad\IEEEauthorrefmark{2}, 
        Muhammad Taimoor Khan\IEEEauthorrefmark{3},\\
        Majid Haghparast\IEEEauthorrefmark{1}, 
        Vlad Stirbu\IEEEauthorrefmark{1}, 
        Peng Liang\IEEEauthorrefmark{4}
    }
    \IEEEauthorblockA{\IEEEauthorrefmark{1}Faculty of Information Technology, University of Jyväskylä, Jyväskylä, Finland}
    \IEEEauthorblockA{\IEEEauthorrefmark{2}School of Computing and Communications, Lancaster University Leipzig, Leipzig, Germany}
    \IEEEauthorblockA{\IEEEauthorrefmark{3}School of Computing and Mathematical Sciences, University of Greenwich, London, UK}
    \IEEEauthorblockA{\IEEEauthorrefmark{4}School of Computer Science, Wuhan University, Wuhan, China}
}

\maketitle

\begin{abstract}
Quantum Software (QSW) aims to exploit the principles of quantum mechanics -- programming quantum bits (qubits) that manipulate quantum gates (qugates) -- to implement quantum computing systems. QSW has emerged as a quantum-specific genre of software-intensive systems, requiring notations, languages, patterns, and tools, etc., for mapping the operations of qubits and the structure of qugates to components and connectors of QSW architectures. To support declarative modelling of QSW, we seek to enable architecture-centric development where software engineers can model, program, and evaluate quantum software systems by abstracting away implementation details through high-level components and connectors. In particular, we propose the specification and present a prototype for QADL -- Quantum Architecture Description Language -- providing a specification language, design space, and execution environment to architect QSW. Inspired by classical ADLs, QADL provides (i) a graphical interface to specify and design QSW components, (ii) a parser for syntactical correctness of the components, and (iii) execution by integrating QADL with IBM Qiskit. The preliminary evaluation of QADL is based on usability analysis by a team of quantum physicists and software engineers, using quantum algorithms such as Quantum Teleportation and Grover’s Search. QADL is a pioneering effort, complementing existing research and development on classical ADLs, to offer a specification language and design environment dedicated to architecting QSW.
A demo is available at \url{https://youtu.be/xaplHH_3NtQ}
\end{abstract}


\input{sections/01_introduction}

\input{sections/02_tool_description.tex}
\input{sections/03_use_cases.tex}
\input{sections/04_limitationandfuturework}
\input{sections/Relatedwork}
\input{sections/05_conclusion}

\bibliographystyle{IEEEtran}
\bibliography{sections/References}

\end{document}

%% file: sections/01_introduction.tex
\section{Introduction}
Quantum computing is growing fast as a powerful technology to solve difficult problems that classical systems cannot handle \cite{horowitz2019quantum, zhao2020quantum}. Quantum systems use superposition, entanglement, and interference to speed up problem-solving in critical areas like cryptography \cite{shor1994algorithms}, optimization~\cite{farhi2014quantum}, machine learning \cite{biamonte2017quantum}, and drug discovery~\cite{cao2019quantum}. Despite its transformative potential, developing QSW (QSW) and hardware remains challenging due to the unique properties of quantum mechanics \cite{de2021materials}. The need for specialized tools to model, design, and execute QSW is becoming increasingly apparent.

\textbf{Problem Statement}: This need was identified through industry collaboration \cite{liimatta2024research}, feedback from interdisciplinary teams focusing on quantum physics, computing, and software engineering, and by conducting a systematic review \cite{khan2023software}, which highlighted the gap between QSW design and its execution across various hardware platforms. While quantum programming languages like Q\# \cite{svore2018q} and frameworks such as Qiskit \cite{qiskit} offer environments for writing and running quantum algorithms, there is a lack of tools for higher-level architectural modeling and visualization of quantum circuits. These frameworks primarily focus on executing algorithms, which leaves a gap when it comes to the design and structural representation of QSW. For quantum engineers and researchers, this lack of visualization and abstraction makes it difficult for them to effectively manage complex quantum circuits, optimize gate operations, and model the interaction between hardware and software components in a cohesive framework.

\textbf{Motivation}: The development of QADL is part of our larger project, \textit{Enhanced Middleware for QSW (EM4QS)}, where our main focus is on enhancing quantum execution environments, creating intermediate representations, and providing a standardized approach to modeling QSW. In collaboration with leading Finnish companies working on quantum computing, including CSC\footnote{CSC: IT Center for Science, providing high-performance computing services in Finland}, Quanscient\footnote{Quanscient: Specializes in quantum computing-based simulation software}, QuantrolOx\footnote{QuantrolOx: Focuses on quantum device control using machine learning techniques}, SemiQon\footnote{SemiQon: Develops semiconductors for quantum computing}, and Unitary Zero Space\footnote{Unitary Zero Space: A quantum computing research organization developing innovative quantum technologies}, this project aims to address several key challenges in quantum computing. The motivation behind QADL is to offer a practical and structured framework for the design, visualization, and modeling of various levels of quantum computing hardware and software, their interconnections, and the associated quantum circuits. The recently published paper \cite{zhao2024towards}, which explores the development of an ADL for hybrid quantum-classical systems, emphasizes the limitations of existing ADLs in handling quantum-specific phenomena such as superposition, entanglement, and decoherence. The paper highlights the integration challenges between quantum and classical components, reinforcing the need for a comprehensive framework that can model, analyze, and optimize interactions\cite{zhao2024towards}. Our work is aligned with this direction, as this QADL have potential to address these  challenges.

Our proposed QADL fills this gap by focusing on low-level circuit representation, enabling users to manage qubits, define gate operations, and map QSW onto hardware, providing a clearer connection between quantum design and execution. QADL differs from both Q\# and Qiskit in its focus and purpose. While Q\# and Qiskit are primarily built for writing and executing quantum algorithms on simulators and real quantum hardware \cite{qiskit}, QADL is designed specifically as a quantum architecture description language. Its emphasis is on structural modeling and visualization at an abstract level, using intuitive syntax for defining circuits. Although QADL currently utilizes Qiskit’s tools for circuit visualization, its primary role is to provide a modular and descriptive framework for defining QSW, bridging the gap between high-level quantum design and low-level execution to achieve the overall goal of the EM4QS project.

\textbf{Contribution}: We introduce QADL, an ADL that solves the shortcomings of current quantum development tools by offering a high-level view of QSW. QADL includes key features like qubit management, gate sequencing, and measurement, which allow for both visualization and modeling of QSW. It can serve as a structured platform for designing quantum circuits, complementing tools like Q\# and Qiskit that focus more on execution of algorithms.

%% file: sections/02_tool_description.tex
\section{Overview}    
\begin{figure}
    \centering
    \includegraphics[width=\linewidth]{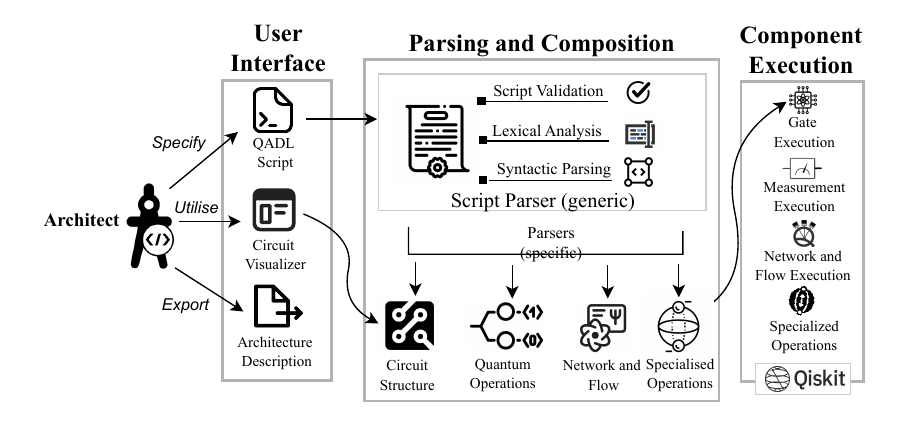}
    \caption{Overview of QADL}
    \label{fig:Architecture}
\end{figure}

    In this paper, we present a prototype of QADL, a specialized ADL for QSW, following an iterative development methodology that aligns with the method proposed by \cite{zhao2024towards}. Through extensive research and analysis, we identified and categorized 50 key quantum system components (e.g., qubits, quantum gates, quantum circuits, quantum buses). We grouped these components into multiple architectural views, including structural, computational, integration, and physical, and defined the syntax and semantics for each component to capture essential quantum properties, such as entanglement and superposition. We developed a prototype of QADL to gather feedback and iteratively improve its usability. Future work includes integrating QADL with existing quantum platforms like Qiskit, enabling practical implementation, and evaluating the framework on real-world quantum architectures. This iterative process will ensure that QADL evolves as a comprehensive tool for designing and describing quantum software architectures. Currently, the prototype of the QADL system consists of three layers: (1) \textit{User Interface}, (2) \textit{Parsing and Composition}, and (3) \textit{Component Execution} (see Figure \ref{fig:Architecture}). We briefly describe each of the QADL components in the following.

\subsection{Graphical User Interface} 
We provide the desktop-based QADL Graphical User Interface (GUI), which gives a user-controlled environment for designing, viewing, and exporting quantum circuits. Below we outline the main features of the GUI.

\begin{itemize} 
\item \textbf{QADL Script Input}: QADL Script defines quantum circuits by structuring and specifying operations in a simple syntax. The user declares a circuit using the keyword \texttt{Circuit}, followed by the name of the circuit and a block enclosed by curly braces \{ \}. Inside this block, the user defines qubits by using the \texttt{qubit} keyword and then applies the gates using the \texttt{gate} keyword, specifying for which qubit the gate operates. The listing of the script is enclosed between the tags \texttt{@startqadl} and \texttt{@endqadl}, that mark the beginning, respectively the end of the QADL code.

\item \textbf{Real-Time Circuit Visualization}: The right-hand panel of the GUI is dynamically updated to visualize the quantum circuit. When writing or editing the script and executing the program from the icon menu, it instantly generates a visual representation of the quantum circuit. It allows users to understand how qubits and gates are set, ordered, and connected at run time.

\item \textbf{Exporting Architecture Description}: Once a circuit is specified and visualized, the system enables users to export the architecture description. The exported file contains all relevant circuit details, including qubit definitions, gate operations, and measurements. This functionality facilitates the sharing and further development of the quantum circuit on other platforms or for simulation purposes.
\end{itemize}

\begin{figure}
   \centering
    \includegraphics[width=0.8\linewidth]{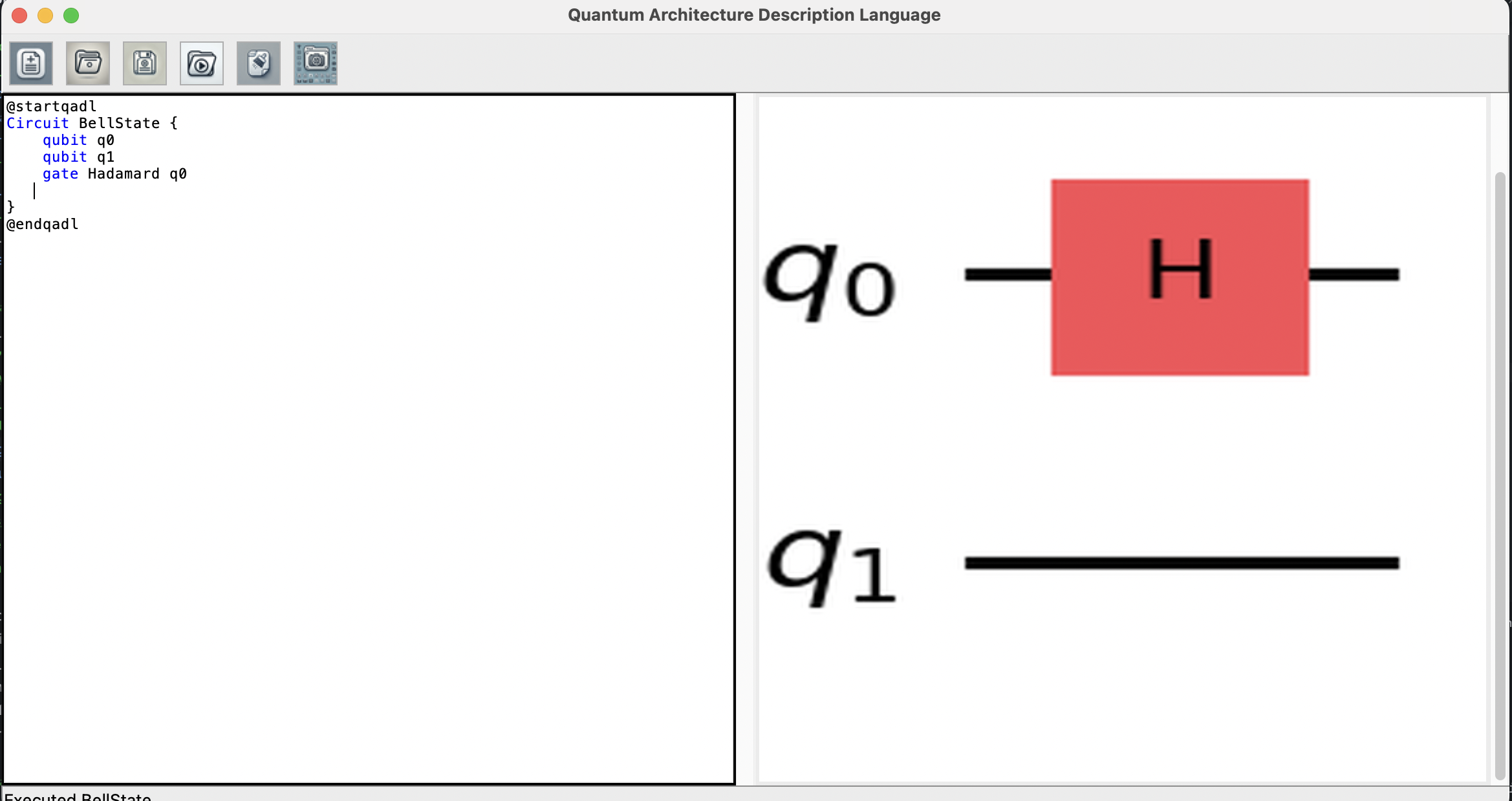}
    \caption{QADL Graphical User Interface }
    \label{fig:GUI}
\end{figure}

\subsection{Parsing}
The QADL parser tokenizes the input script, breaking it down into components like qubits, gate operations, and measurement declarations. While processing each statement, it applies syntactic rules to validate the circuit. In case of a syntax error, feedback is provided, otherwise, the parser constructs a quantum circuit representation by identifying key components and ensuring that each element follows the defined syntax and structure. The QADL parser is structured into several sub-parsers to efficiently handle quantum circuit elements and support future evolution. These sub-parsers are presented as follows:

\begin{itemize}
    \item \textbf{Circuit Structure Parser}: This ranges from modules responsible for managing the overall structure in the QADL script to handling circuit declarations and parsing blocks of quantum operations, including validating the correct flow within the circuit. Most of the above-mentioned functionality is achieved through dedicated parsing mechanisms.
    \item \textbf{Quantum Operations Parser}: This module handles parsing for quantum operations, e.g., applying gates like \texttt{Hadamard, CNOT}, and \texttt{CRZ}, as well as measurement operations, which convert quantum states to classical bits. 
   \item \textbf{Network and Flow Parser}: This module is responsible for parsing nodes, edges, and conditional flows to construct quantum circuits, either through dynamic routing or controlled gate operations based on measurement results.
   \item \textbf{Specialized Parsers}: These module consolidate the parsed elements to form a coherent but valid quantum circuit. Modules like \texttt{qadl3\_parser} coordinate the parser's constituent parts so that qubits are correctly connected with their gates and measurements in the circuit representation.
\end{itemize}

\subsection{Component Execution}

After successfully parsing the QADL script, the next step is to execute the quantum circuit. In this phase, the parsed elements (e.g., qubits, gates, measurements) are translated into a functional quantum circuit using Qiskit. The execution of each statement is integrated with the QADL GUI and parser to ensure a smooth and efficient translation of script input into circuit simulation.

\begin{itemize}
    \item \textbf{Gate Execution}: The parsed gates (e.g., Hadamard, CNOT, CZ, CRZ) are applied to the qubits as specified in the QADL script. For example, a Hadamard gate is applied to a qubit by calling the corresponding Qiskit operation. Each gate should be correctly mapped to its quantum equivalent and implemented accurately.
    \item \textbf{Measurement Execution}: This module is responsible for extracting measurements from the script, performing operations on qubits, and storing the results in classical bits. We also provide flexibility, allowing measurements to use conditional and iteration (e.g., \texttt{if} or \texttt{for}) statements to process the output of quantum computation results.
    \item \textbf{Network and Flow Execution:} We provide features for network and execution flow in QADL to handle more complex circuits involving dynamic routing, conditional operations, nodes, edges, and flows. These features are essential for circuits that require interactions based on measurement outcomes.
    \item \textbf{Specialized Operations}: This feature handles advanced operations (e.g., the inverse Quantum Fourier Transform (QFT) and Grover’s Oracle) through dedicated functions such as \texttt{apply\_inverse\_qft} and \texttt{grovers\_oracle}. These specialized functions simplify the execution of complex quantum algorithms.
\end{itemize}

%% file: sections/03_use_cases.tex
\section{Use Case and Early Validation}
To validate the capabilities of QADL, we conducted tests for well-known quantum algorithms. Due to limited space, we only provide the mathematical form -- QADL description -- and visualization of Quantum Teleportation \cite{} and Grover's search in Section \ref{Grover}. The primary purpose of this use case is to demonstrate how complex mathematical quantum equations can be described and visualized using QADL script.

\subsection{Quantum Teleportation}
\label{Teleportation}
\subsubsection{Mathematical Form}
Quantum teleportation is primarily used for transferring quantum information between two locations without physically moving qubits, by utilizing a combination of quantum entanglement and classical communication. In the following, we briefly present and explain the mathematical form of the quantum teleportation algorithm

Quantum teleportation transfers quantum information between two locations using quantum entanglement and classical communication. We begin with three qubits: Qubit \( q_1 \) holds the unknown quantum state to be teleported:

\[
\vert \psi \rangle = \alpha \vert 0 \rangle + \beta \vert 1 \rangle
\]

Qubits \( q_2 \) and \( q_3 \) form an entangled pair in the Bell state:

\[
\vert \Phi^+ \rangle = \frac{1}{\sqrt{2}} (\vert 00 \rangle + \vert 11 \rangle)
\]

After applying a Hadamard gate to \( q_1 \) and a CNOT gate between \( q_1 \) and \( q_2 \), the total state becomes:

\[
\frac{1}{2} (\alpha \vert 000 \rangle + \alpha \vert 011 \rangle + \beta \vert 100 \rangle + \beta \vert 111 \rangle)
\]

The measurement result is sent to the receiver via classical communication. The receiver then applies one of the following operations on \( q_3 \), depending on the classical bits:

- No correction for \( \vert 00 \rangle \)
- Apply Pauli-X for \( \vert 01 \rangle \):
\[
X = \begin{pmatrix} 0 & 1 \\ 1 & 0 \end{pmatrix}
\]
- Apply Pauli-Z for \( \vert 10 \rangle \):
\[
Z = \begin{pmatrix} 1 & 0 \\ 0 & -1 \end{pmatrix}
\]
- Apply Pauli-XZ for \( \vert 11 \rangle \)

After the correction, the final state of \( q_3 \) is:

\[
\vert \psi \rangle = \alpha \vert 0 \rangle + \beta \vert 1 \rangle
\]

\begin{figure}[h]
    \caption{Quantum Teleportation Algorithm: The steps for teleporting the state of qubit \( q_1 \) to qubit \( q_3 \) using an entangled pair and classical communication.}
\end{figure}

\begin{equation}
\vert \psi \rangle = \alpha \vert 0 \rangle + \beta \vert 1 \rangle
\end{equation}
Qubits \( q_2 \) and \( q_3 \) form an entangled pair in the Bell state:
\begin{equation}
\vert \Phi^+ \rangle = \frac{1}{\sqrt{2}} (\vert 00 \rangle + \vert 11 \rangle)
\end{equation}
The total initial state of the three qubits is:
\begin{equation}
\vert \psi \rangle \otimes \vert \Phi^+ \rangle = (\alpha \vert 0 \rangle + \beta \vert 1 \rangle) \otimes \frac{1}{\sqrt{2}} (\vert 00 \rangle + \vert 11 \rangle)
\end{equation}
A Hadamard gate is applied to \( q_1 \), and a controlled-NOT (CNOT) gate is applied between \( q_1 \) and \( q_2 \), preparing them for Bell measurement:
\begin{equation}
H(q_1), \quad CNOT(q_1, q_2)
\end{equation}
The sender performs a Bell-state measurement on \( q_1 \) and \( q_2 \). After the CNOT gate and a Hadamard gate on \( q_1 \), the state becomes:
\begin{equation}
\frac{1}{2} \left( \alpha \vert 000 \rangle + \alpha \vert 011 \rangle + \beta \vert 100 \rangle + \beta \vert 111 \rangle \right)
\end{equation}
The measurement result is transmitted to the receiver via classical communication. The sender measures \( q_1 \) and \( q_2 \), producing two classical bits, which are sent to the receiver.

Based on the classical bits, the receiver applies one of the following unitary operations on qubit \( q_3 \):
No correction for \( \vert 00 \rangle \), apply Pauli-X for \( \vert 01 \rangle \):
\begin{equation}
X = \begin{pmatrix} 0 & 1 \\ 1 & 0 \end{pmatrix}
\end{equation}
Apply Pauli-Z for \( \vert 10 \rangle \):
\begin{equation}
Z = \begin{pmatrix} 1 & 0 \\ 0 & -1 \end{pmatrix}
\end{equation}
or apply Pauli-XZ for \( \vert 11 \rangle \).

After the correction, the final state of \( q_3 \) is:
\begin{equation}
\vert \psi \rangle = \alpha \vert 0 \rangle + \beta \vert 1 \rangle
\end{equation}

\begin{figure}[h]
    \caption{Quantum Teleportation Algorithm: This algorithm describes the steps required to teleport the state of qubit \( q_1 \) to qubit \( q_3 \) using an entangled pair and classical communication.}
\end{figure}

\subsubsection{QADL Description and Visualization}
We defined the script of the QADL for quantum teleportation, as shown in Figure \ref{QuantumTeleportation}. It initializes three qubits: \( q0 \), \( q1 \), and \( q2 \). A Hadamard gate is applied to \( q1 \) to create superposition, followed by a CNOT gate between \( q1 \) and \( q2 \) to establish entanglement. Then, a CNOT gate and a Hadamard gate are applied between \( q0 \) and \( q1 \) to prepare them for measurement. The classical bits \( c0 \) and \( c1 \), resulting from the measurements of \( q0 \) and \( q1 \), are used to conditionally apply Pauli-Z and Pauli-X gates to \( q2 \), completing the teleportation process.

\begin{figure}
\begin{lstlisting}
@startqadl
Circuit QuantumTeleportation {

    // Define qubits
    qubit q0
    qubit q1
    qubit q2

    // Step 1: Entanglement
    gate Hadamard q1
    gate CNOT q1 q2

    // Step 2: Bell-state measurement
    gate CNOT q0 q1
    gate Hadamard q0

    // Step 3: Measure and store results
    measure q0 -> c0
    measure q1 -> c1

    // Step 4: Conditional operations
    gate CNOT q1 q2
    gate CZ q0 q2

    // Step 5: Measure and store results
    measure q2 -> c2
}
@endqadl
\end{lstlisting}
\label{QuantumTeleportation}
\caption{QADL Script for Quantum Teleportation}
\end{figure}

\begin{figure}
    \centering
    \includegraphics[width=\linewidth]{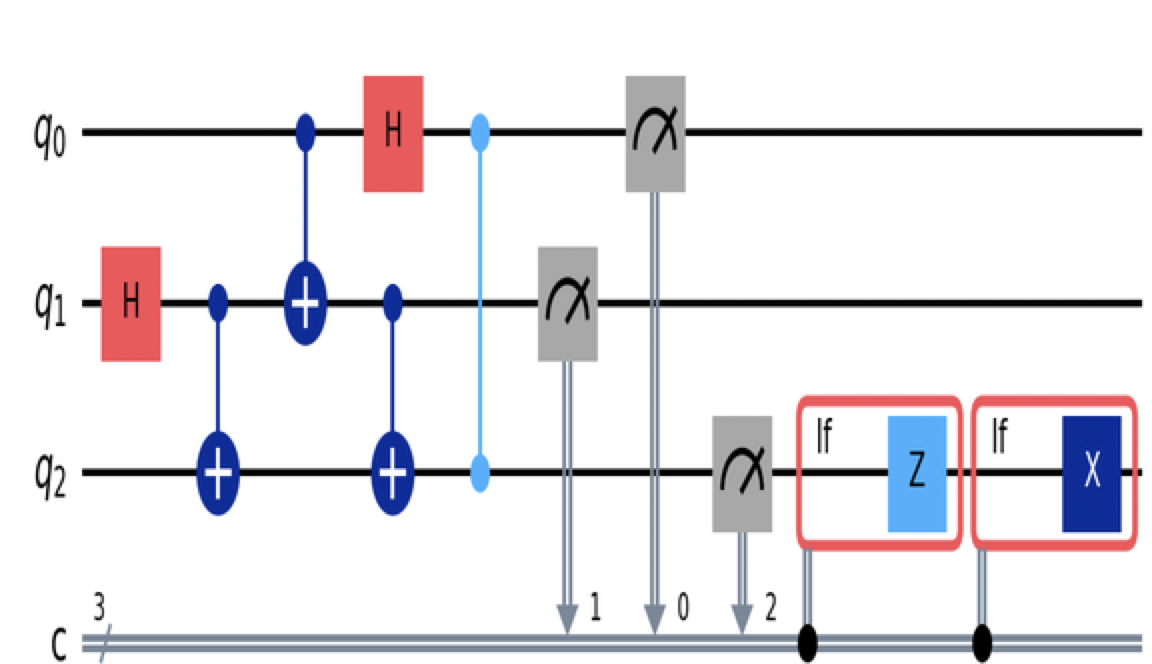}
    \caption{Quantum Teleportation Circuit Generated from the QADL Script}
    \label{fig:teleportation}
\end{figure}

Once we executed the quantum teleportation script, we obtained Figure \ref{fig:teleportation}, which shows the quantum circuit. The Hadamard and CNOT gates create superposition and entanglement, and the measurements of \( q0 \) and \( q1 \) determine whether Pauli-Z and Pauli-X gates are applied to \( q2 \), ensuring the quantum state is teleported from \( q0 \) to \( q2 \) based on the classical results \( c_0 \) and \( c_1 \).

\subsection{Grover's Search Algorithm}
\label{Grover}
\subsubsection{Mathematical Form}
Grover's Algorithm \cite{grover1996fast} searches an unstructured database with a quadratic speedup, and is mathematically expressed as follows:
First, during the initialization phase, Hadamard gates create a superposition of all \( N = 2^n \) possible states.
The initial state is constructed using a Hadamard gate on \( n \) qubits:

\[
H^{\otimes n} \vert 0 \rangle^{\otimes n} = \frac{1}{\sqrt{N}} \sum_{x=0}^{N-1} \vert x \rangle
\]

The oracle \( O \) marks the target state \( s \) by flipping its amplitude:

\[
O \vert x \rangle =
\begin{cases} 
-\vert x \rangle & \text{if } x = s \\
\vert x \rangle & \text{if } x \neq s
\end{cases}
\]

The diffusion operator amplifies the marked state's amplitude:

\[
D = 2 \vert \psi \rangle \langle \psi \vert - I
\]

After applying the oracle and diffusion operator \( k \) times, the state changes into:

\[
\vert \psi_k \rangle = (DO)^k H^{\otimes n} \vert 0 \rangle^{\otimes n}
\]

The number of iterations is approximately \( k \approx \frac{\pi}{4} \sqrt{N} \).

Finally, measuring the system yields the target state with high probability.

\begin{equation}
H^{\otimes n} \vert 0 \rangle^{\otimes n} = \frac{1}{\sqrt{N}} \sum_{x=0}^{N-1} \vert x \rangle
\end{equation}

Next, the oracle \( O \) marks the target state \( s \) by flipping its amplitude:

\begin{equation}
O \vert x \rangle =
\begin{cases} 
-\vert x \rangle & \text{if } x = s \\
\vert x \rangle & \text{if } x \neq s
\end{cases}
\end{equation}

The diffusion operator amplifies the amplitude of the marked state:

\begin{equation}
D = 2 \vert \psi \rangle \langle \psi \vert - I
\end{equation}

where \( \vert \psi \rangle \) is the equal superposition state.

After applying the oracle and diffusion operator \( k \) times, the state evolves as follows:

\begin{equation}
\vert \psi_k \rangle = (DO)^k H^{\otimes n} \vert 0 \rangle^{\otimes n}
\end{equation}

The number of iterations \( k \) is approximately:

\begin{equation}
k \approx \frac{\pi}{4} \sqrt{N}
\end{equation}

Finally, after these iterations, measuring the system yields the target state with high probability.

\subsubsection{QADL Description and Visualization}
We define the script of QADL for Grover’s Algorithm as shown in Figure~3. This script initializes three qubits (\texttt{q0}, \texttt{q1}, \texttt{q2}) by applying Hadamard gates to place them into superposition. After the initialization, an Oracle is applied to mark the desired solution. This is followed by the Diffusion operator (Grover diffusion), which inverts the amplitudes around the average to amplify the correct state. Finally, all three qubits are measured, and the results are stored in classical bits (\texttt{c0}, \texttt{c1}, \texttt{c2}). The QADL script reflects the structure of Grover's search algorithm, which is designed to search for a marked item in an unsorted database efficiently.

\begin{figure}
\scriptsize
\label{fig:GroversAlgorithm}
\begin{lstlisting}
@startqadl
Circuit GroversAlgorithm {
    qubit q0, q1, q2
    gate Hadamard q0, q1, q2
    gate X q0   // Invert q0
    gate X q2   // Invert q2
    gate CNOT q0 q1  // Controlled q0, q1
    gate CNOT q1 q2  // Controlled q1, q2
    gate X q0   // Undo inversion of q0
    gate X q2   // Undo inversion of q2
    gate Hadamard q0, q1, q2
    gate X q0, q1, q2
    gate CNOT q0 q1  
    gate CNOT q1 q2
    gate X q0, q1, q2
    gate Hadamard q0, q1, q2
    measure q0 -> c0
    measure q1 -> c1
    measure q2 -> c2
}
@endqadl
\end{lstlisting}

\caption{QADL Script for Grover's Algorithm}
\end{figure}
\begin{figure}[h!]
    \centering
    \includegraphics[width=\linewidth]{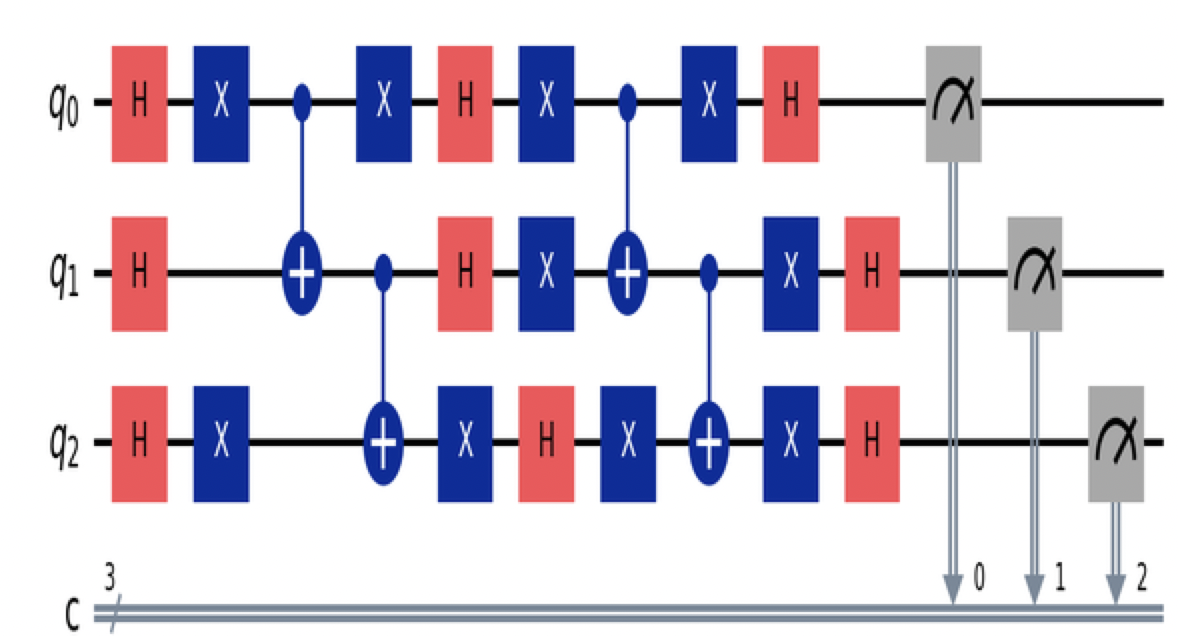}
    \caption{Grover's Algorithm Circuit Generated from the QADL Script}
    \label{fig:grover_circuit}
\end{figure}

The script is parsed and executed through a quantum circuit simulator, where each gate is translated into its equivalent Qiskit function. After execution, a visual representation of the circuit can be generated. Figure~\ref{fig:grover_circuit} shows the resulting quantum circuit, visualized with the qubits and gates as specified in the QADL script.

\subsection{Early Validation}
This prototype of QADL was tested within our research group, which contained a diverse set of experts in quantum physics and quantum software engineering. The test included several iterative feedback cycles where users from diverse backgrounds interacted with QADL by designing and visualizing quantum circuits, including famous algorithms such as Quantum Teleportation and Grover's Search. All of them provided invaluable feedback regarding the usability of the tool, the clarity of its syntax, and the error-handling mechanisms that helped us develop the improved script. Furthermore, practical evaluation ensured that this prototype is user-friendly and capable of modeling complex quantum circuits, meeting both novice and expert users' needs in quantum computing.

%% file: sections/04_limitationandfuturework.tex
\section{Limitation and Future Work}
\label{limitationandfuturework}
The current QADL prototype depends on Qiskit for visualization and execution, which limits its compatibility among other quantum platforms like Cirq or ProjectQ. In the future, we plan to expand support to these platforms to enhance its flexibility. Furthermore, QADL mainly addresses low to intermediate-level circuit descriptions and lacks advanced capabilities such as error correction, noise modeling, and hardware-specific optimizations. Adding these features will be a key focus in the next phase to enable QADL to support fault-tolerant quantum computing systems. Moreover, while internal testing has provided insights, QADL still needs extensive validation with external industry partners. We aim to collaborate with these partners to ensure the tool's practical relevance. Lastly, the tool currently lacks advanced debugging features, such as quantum state tracking and error diagnostics, which will be prioritized in the next development cycle.

Our roadmap includes expanding platform compatibility, introducing advanced features, and enhancing debugging and testing to meet broader industrial requirements. To further enhance validity of the tool, we are formalising semantics~\cite{omni} of the QADL constructs to assure that the tool only accepts semantically sound quantum circuits and their interactions, i.e., mapping a valid input (quantum) state to a set of valid output (quantum) states considering the effect of superposition and entanglements.

%% file: sections/Relatedwork.tex
\section{Related Work}\label{Related Work}

Quantum programming languages like Q\# from Microsoft and Qiskit from IBM already provided strong templates in developing and executing quantum algorithms. Q\# includes abstractions at a high level to make quantum programming more accessible, whereas Qiskit allows the execution of algorithms on simulated and real quantum hardware \cite{microsoftQSharp, qiskit}. In contrast, neither of the languages specifically has tools for architectural descriptions, intuitive modeling, or visualization from the structural point of view w.r.t. QSW. This has been a gap in the tight management that a quantum engineer needs to perform at the architectural level. Several packages such as ProjectQ \cite{projectq}, Cirq \cite{cirq}, and OpenQL \cite{openql} have attempted to provide visualizations and modular circuit composition, but they mainly focus on gate-level design rather than a comprehensive architectural description. To our knowledge, none provide an intermediate-level quantum architecture description that integrates design, execution and visualization. To this end, QADL offers a modular and mid-level language that defines qubits, gates, and measurements. Different from other tools, QADL integrates Qiskit for smooth visualization and execution.

%% file: sections/05_conclusion.tex
\section{Conclusions}

In this paper, we present QADL, a language designed to simplify the modeling, execution, and visualization of quantum architectures. Through its user interface, parser, and execution components, QADL provides a structured framework for designing and visualizing quantum circuits. Early testing with quantum algorithms such as Quantum Teleportation and Grover's Search demonstrated its functionality and potential for broader applications. 